\documentclass{ws-procs975x65}
\usepackage{graphicx}

\begin{document}

\title{Weak Chaos in large conservative system -- Infinite-range coupled standard maps}

\author{Luis G. Moyano}

\address{Centro Brasileiro de Pesquisas F\'{\i}sicas\\
Rua Xavier Sigaud 150, Urca 22290-180\\
Rio de Janeiro, Brazil\\
E-mail: moyano@cbpf.br}

\author{Ana  P. Majtey}

\address{Facultad de Matem\'atica, Astronom\'{\i}a y F\'{\i}sica\\
Universidad Nacional de C\'ordoba, Ciudad Universitaria\\
5000 C\'ordoba, Argentina\\
E-mail: amajtey@famaf.unc.edu.ar}  

\author{Constantino Tsallis}

\address{Santa Fe Institute,\\
1399 Hyde Park Road, 
Santa Fe, NM 87501, USA\\
E-mail: tsallis@santafe.edu\\
and\\
Centro Brasileiro de Pesquisas F\'{\i}sicas,\\
Rua Xavier Sigaud 150, Urca 22290-180,\\
Rio de Janeiro, Brazil}
 
\maketitle

\abstracts{
We study,  through a new perspective, a globally coupled map system that essentially interpolates between simple discrete-time 
nonlinear dynamics and certain long-range many-body Hamiltonian models. In particular, we exhibit relevant similarities, namely  (i) the existence of long-standing quasistationary states (QSS), and (ii) the emergence of weak chaos in the thermodynamic limit, between the present model and the Hamiltonian Mean Field model, a strong candidate  for a nonxtensive statistical mechanical approach. 
}


\section{Introduction}

In the last years, considerable effort has been made in order to clarify the role that 
nonextensive statistical mechanics\cite{tsallis1} plays in physics. In this context, there has been 
significantly growing evidence relating several, physically motivated, nonlinear dynamical systems.
It has been repeatedly put forward  that the statistical behaviour of a physical system 
descends from its microscopic dynamics\cite{tsallis2}. Consequently, the study of paradigmatic nonlinear dynamical systems is important  in order to describe and understand anomalies and deviations from the well known Boltzmann-Gibbs (BG) statistical mechanics.
The scenario within which we are working tries to capture the most relevant features of nonextensive statistical mechanics in the 
complete range of dynamical systems: from extremely simple {\it dissipative} low-dimensional maps to complex {\it conservative} 
Hamiltonian dynamics\cite{tsallis3}. In the present paper we will make specific progress along these lines by focusing on a model which illustrates the deep similarities that can exist between nonlinear coupled maps and many-body Hamiltonian dynamics.


Let us first recall a paradigmatic and intensively studied many-body infinite-range coupled conservative system, namely
the Hamiltonian Mean Field (HMF) model\cite{antoni,dauxois}:

\begin{equation}
H = K+V=\sum_{i=1} ^{N} \frac{p_{i}^2}{2} +
\frac{1}{2N} \sum_{i,j=1} ^{N}
\left[1-cos(\theta_{i}-\theta_{j})\right].
\label{HMF}
\end{equation}
The HMF model may be thought of as $N$ globally coupled classical spins (inertial version of 
the $XY$ ferromagnetic model). Its molecular dynamics exhibit a remarkably rich behaviour. When the 
initial conditions are out of equilibrium (for example, the so called {\em waterbag}  initial conditions\cite{pluchino1}), 
it can present an
anomalous {\it temperature} evolution (we consider $T \equiv 2K/N$, being $K$ the total kinetic energy). 
These states are characterized by a first stage ({\it quasistationary state}, QSS)
whose temperature is different from that predicted by the BG theory, 
followed by a {\it crossover} to the expected final temperature. These QSS appear to be a consequence of the long-range coupling. They are important because their duration 
diverges with $N$, thus becoming the only relevant state for a macroscopic system\cite{pluchino2}.

At the other end of the range of dynamical systems we may consider  
a dissipative, one-dimensional model, such as the logistic map (and its universality class):  
\begin{equation}
x_{t+1}=1-\mu \, x_t^2   \;\;\;\;\;\;
(t=0,1,2...; \,
x\in[-1,1]; \,
\mu\in[0,2])\,.
\label{logistic}
\end{equation}
Because of its physical importance, the logistic map is one of the most studied low-dimensional maps. Despite its apparently simple form, it exhibits a 
quite complex behaviour. Important progress has recently been made which places this model as an important example of the applicability of nonextensive statistical mechanical concepts. Indeed, Baldovin and Robledo\cite{fulvio3} rigorously proved that, at the edge of chaos (as well as at the doubling-period and tangent bifucations), the sensitivity to 
initial conditions is given by a $q$-exponential function \cite{tsallis1}. Furthermore, they proved\cite{fulvio3} the $q$-generalization of a Pesin-like identity concerning the entropy production per unit time. For the stationary state at the edge of chaos, the entropic
index $q$ can be obtained analytically. 
Moreover, when a small external noise is added, a two-step relaxation evolution is found\cite{fulvio2}, similarly to what occurs for the HMF case. 

At this point, a natural question may arise. Is it possible to 
relate the results found for such simple maps to the anomalies found in the
HMF model? Furthermore, can we treat various nonlinear dynamical systems within the nonextensive
statistical mechanics theory? Many studies are presently addressing such questions.

A first step that can be done in this direction is to move closer to a Hamiltonian dynamics by considering a symplectic, 
conservative map. This is the case of the widely investigated Taylor-Chirikov {\em standard map}\cite{ott}:
\begin{eqnarray}
\begin{array}{rrcl}
\theta(t+1)&
=&
p(t+1) + \theta(t)\hspace{.9cm}&
\hspace{1cm} \text{(mod\,1)},\\
p(t+1)&
=&
p(t) + \frac{a}{2\pi}\sin[2\pi\theta(t)]&
\hspace{1cm} \text{(mod\,1)}\,.
\end{array}
\label{standard}
\end{eqnarray}
This map may be obtained, for instance,  
by approximating the differential equation of a simple pendulum by a centered difference equation, and converting a 
second-order equation into two first-order equations. 

The standard map was studied along the present lines by Baldovin {\em et al}\cite{fulvio1}. For symplectic maps,
what plays a role analogous to the temperature
is the variance of the angular momentum: $T\equiv \sigma^2_p = \langle p^2 \rangle - 
\langle p \rangle ^2$, where $\langle\, \rangle$ denotes the ensemble average. 
Beginning with the same type of initial  conditions as before ({\em waterbag}), 
we observe once again a two-plateaux relaxation process, suggesting a
connection with the phenomena already described for the HMF model.

A step forward to capture the behaviour of the HMF system of rotors 
is to consider $N$ standard maps, coupled in such way as to mantain their symplectic (hence conservative) structure. 
However, there are several ways to achieve this. A particular coupling {\em in the momenta} has been 
recently addressed\cite{fulvio2} with quite interesting results such as QSS relaxation and 
nonergodic occupation of phase space. A different type of coupling is addressed in the next Section.
\section{Symplectic coupling in the coordinates}  
As before, we consider $N$ standard maps 
but, this time, with a global, symplectic coupling {\em in the coordinates}\,:
\begin{eqnarray}
\begin{array}{rclr}
\theta_i(t+1)&
=&
\theta_i(t) + p_i(t+1),&
\hspace{.3cm} \text{(mod\,1)}\,,\\ 
p_i(t+1)&
=&
p_i(t) + \frac{a}{2\pi}\sin[2\pi\theta_i(t)]+\frac{b}{2\pi(N-1)}\sum_{j\not=i}^N \sin[2\pi(\theta_i(t)-\theta_j(t))]&
\hspace{.5cm} \text{(mod\,1)}\,.
\end{array}
\label{AZP}
\end{eqnarray}
This coupling arises as a natural choice. In fact, applying to the 
HMF model the difference procedure mentioned  above for the standard map, we obtain precisely the $a=0$ particular instance of model (4). 
This model has already been addressed in the literature\cite{azp}, 
but in a quite different context, related to the study of the Lyapunov exponents in the completely chaotic regime. 

\begin{figure}[h!] 
\centering 
 \includegraphics[width=0.9\textwidth]{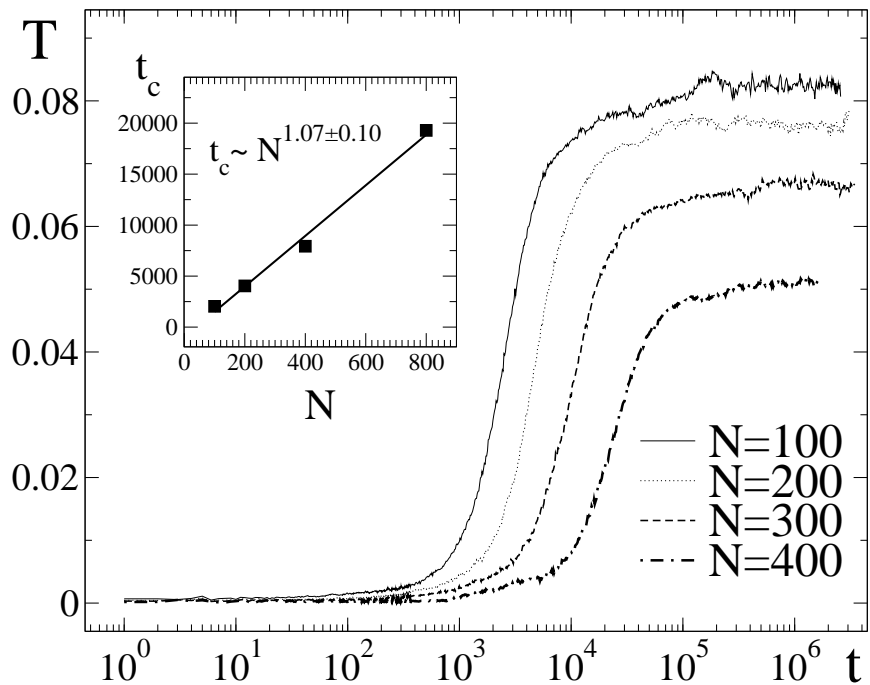}
\caption{{\it Temperature} evolution illustrating the presence of two-step
  relaxation (QSS) for typical system sizes. We have used $a=0.05$, $b=2$, and
  waterbag initial conditions. Ensemble averages were done,
typically over $100$ realizations. Only much longer simulations could confirm, or exclude, the possibility that all curves, i.e. $\forall N$, saturate at the {\it equal-probability} value $1/12 \simeq 0.08$. 
{\it Inset:} The crossover time $t_c$ corresponds to the inflexion point of $T$ {\it versus} $\log t$.\label{fig1}}
\end{figure}

We present next numerical simulations of the map system (4). We calculated 
the evolution of the variance of the momenta $\sigma^2_p$. Our results show that, for {\em waterbag} 
initial conditions and appropriate ranges for the parameters $a$ and $b$, two-step relaxation processes are 
again found. In Fig. \ref{fig1} we show these results for different sizes of the system. It can be seen that 
the crossover time $t_c$ grows as $t_c\sim N^{1.07\pm 0.10}$, thus never reaching BG equilibrium 
when $N \rightarrow \infty$. In other words, the $N \to\infty$ and $t \to\infty$ limits do not commute. 

Finally, we calculated the {\it largest Lyapunov exponent} (LLE) $\lambda_L$ (we recall that Lyapunov exponents measure the instability of 
dynamical trajectories, and provide a quantitative idea of the sensitivity to the initial conditions of the system). 
Indeed, for the $(a,b)$-parameters in the range illustrated in Figs. 1 and 2, we found that the dependence 
of the LLE with the sistem size is consistent with $\lambda_L \sim N^{-0.40\pm 0.08}$, i.e., a clear indication of {\em weak chaos} 
in the thermodynamic limit.      

\begin{figure}[h!] 
\centering 
\includegraphics[width=0.9\textwidth]{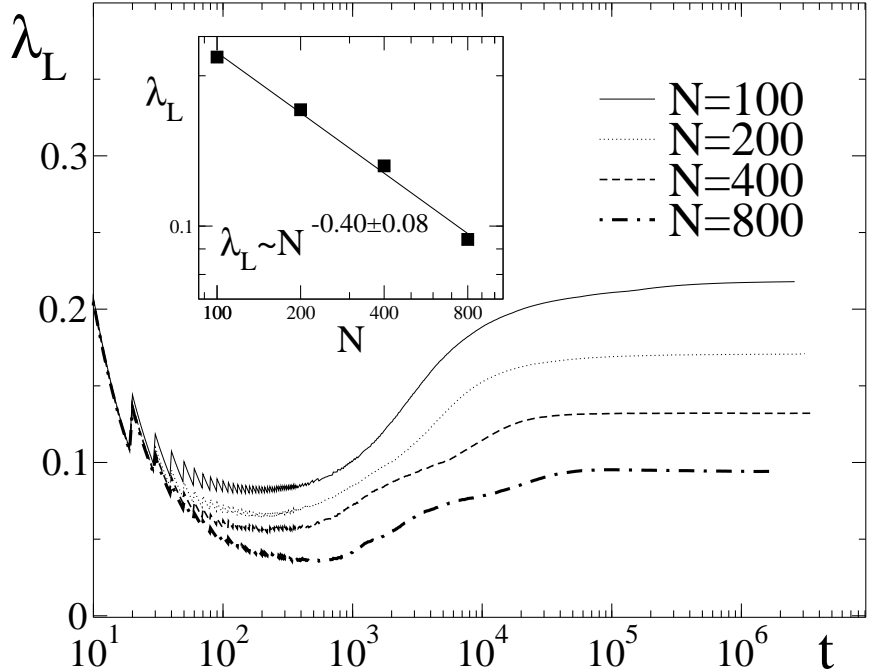}
\caption{Time dependence of the {\it effective} largest Lyapunov exponent $\lambda_L$ for typical  sizes (same parameters as in Fig. 1). 
{\it Inset:} $N$-dependence of the asymptotic value of $\lambda_L$, consistent  with weak chaos in the thermodynamic limit.\label{fig2}}
\end{figure}

Summarizing, we presented a conservative model consisting in $N$ standard maps symplectically coupled through 
the coordinates. We have found results suggestively similar to those obtained for other nonlinear dynamical systems 
including the HMF model. More specifically, we found the double plateaux in the time evolution of the temperature, and a LLE which approaches zero for increasing size. We are currently studying several other quantities (e.g., correlation functions and momenta probability distribution functions), as well as the influence of $(a,b)$ on the present ones. These results place naturally the present system within a series of nonlinear dynamical systems which starts with one-dimensional dissipative maps, follows with low-dimensional conservative maps, then many symplectically coupled maps, and ends with long-range many-body Hamiltonians. They all share important phenomena, typically related, in one way or another, to weak chaos and long-standing nonergodic occupation of phase space. These features precisely constitute the scenario within which nonextensive statistical mechanics appears to be the adequate thermostatistical theory, in analogy to Boltzmann-Gibbs statistical mechanicis, successfully used  since more than one century for strongly chaotic and ergodic systems.

LGM  thanks the organizers for warm hospitality at the meeting in Erice, Italy, 
in particular A. Rapisarda. We would like to thank Stefan Thurner for his useful remarks. Partial finantial support from CNPq, Faperj and Pronex/MCT (Brazilian agencies) is acknowledged as well.

\end{document}